\documentclass{article}

\usepackage{arxiv}

\usepackage{cite}
\usepackage{amsmath,amssymb,amsfonts}
\usepackage{algorithmic}
\usepackage{graphicx}
\usepackage{textcomp}
\usepackage{xcolor}

\usepackage{doi}

\def\BibTeX{{\rm B\kern-.05em{\sc i\kern-.025em b}\kern-.08em
    T\kern-.1667em\lower.7ex\hbox{E}\kern-.125emX}}

\author{
 Reiji Suzuki \\
  Graduate School of Informatics\\
  Nagoya University\\
  Furo-cho, Chikusa-ku, Nagoya 464-8601, Japan \\
  \texttt{reiji@nagoya-u.jp} \\
   \And
 Takaya Arita \\
  Graduate School of Informatics\\
  Nagoya University\\
  Furo-cho, Chikusa-ku, Nagoya 464-8601, Japan \\
  \texttt{arita@nagoya-u.jp} \\
}

\begin{document}

\title{Evolutionary ecology of words\\
}

\maketitle

\begin{abstract}
We propose a model for the evolutionary ecology of words as one attempt to extend evolutionary game theory and agent-based models by utilizing the rich linguistic expressions of Large Language Models (LLMs). Our model enables the emergence and evolution of diverse and infinite options for interactions among agents. Within the population, each agent possesses a short word (or phrase) generated by an LLM and moves within a spatial environment. When agents become adjacent, the outcome of their interaction is determined by the LLM based on the relationship between their words, with the loser's word being replaced by the winner's. Word mutations, also based on LLM outputs, may occur. We conducted preliminary experiments assuming that ``strong animal species" would survive. The results showed that from an initial population consisting of well-known species, many species emerged both gradually and in a punctuated equilibrium manner. Each trial demonstrated the unique evolution of diverse populations, with one type of large species becoming dominant, such as terrestrial animals, marine life, or extinct species, which were ecologically specialized and adapted ones across diverse extreme habitats. We also conducted a long-term experiment with a large population, demonstrating the emergence and coexistence of diverse species.

\end{abstract}


\section{Introduction}

Recent progress in large language models (LLMs) has made it possible to incorporate real-world levels of complexity and diversity into computational models using natural language, leading to various applications in complex systems science \cite{Lu:2024} and artificial life research \cite{Nisioti:2024}. 

In particular, the use of LLMs as autonomous agents has been discussed in machine psychology \cite{Hagendorff:2023}, which involves understanding the cognitive functions of LLMs and using them as virtual human subjects. LLM-based multi-agent research has also emerged, which investigates the dynamics of interactions among LLM agents and applies them to collective problem-solving \cite{Guo:2024,Gao:2024}. 

There is also research focusing on utilizing LLMs to generate novel and diverse linguistic expressions that reflect given contexts. Applications of LLMs to evolutionary computation have been explored \cite{Wu:2024}. Meyerson et al. proposed a language model crossover based on few-shot prompting, which inputs several patterns as parents to an LLM to generate their offspring patterns \cite{Meyerson:2023}. They have successfully evolved binary bit strings, sentences, equations, text-to-image prompts, and Python code. Fernando et al. also proposed a framework where the mutation method described in prompts to LLMs is itself held by organisms as genes and evolves \cite{Fernando:2023}. Lim et al. demonstrated the usefulness of providing information on the characteristics and adaptivity of solutions to LLMs for few-shot learning to generate new solutions in MAP-Elites, a Quality-Diversity algorithm that emphasizes the balance between solution diversity and quality \cite{Lim:2024}. Furthermore, there are studies on evolutionary search in the latent space of generative models \cite{Machin2022} and an approach that utilizes LLMs to improve the Differential Evolution algorithm itself \cite{Pluhacek:2024}. 

Recently, there have been several approaches related to both aspects of the use of LLMs. Suzuki et al. demonstrated the emergence of diverse personality traits or personas in the evolution of LLM agent populations playing the Prisoner's Dilemma by generating their behaviors based on linguistically described personality traits using LLMs \cite{Suzuki:2024a}. Nisioti et al. also examined open-ended creative activities in LLM multi-agent populations using the game Little Alchemy 2, where objects are combined to create new ones. They demonstrated that dynamic network structures determining references to others' behaviors contribute to the generation of highly creative products \cite{Nisioti:2024b}. 

Therefore, understanding and applying the novel and creative generative capabilities of LLMs is an important theme not only for engineering applications of evolutionary systems but also for open-ended evolution, which is a core concept in artificial life research that explores systems capable of continuous innovation and increasing complexity.

\begin{figure*} % picture
    \centering
    \includegraphics[width=150mm]{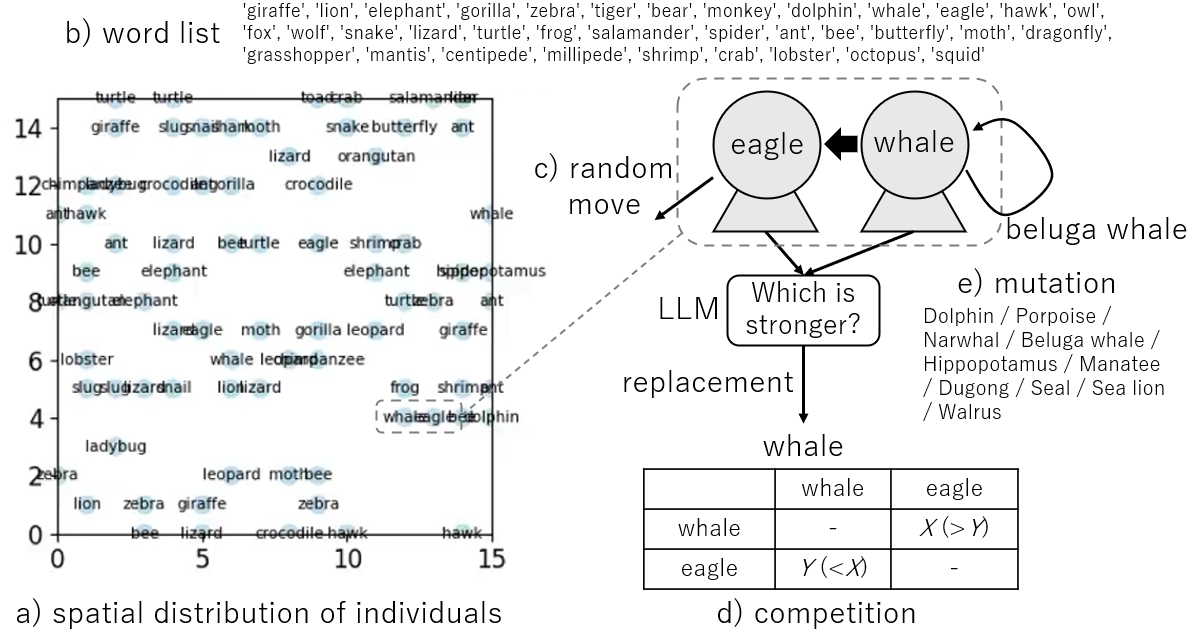}
  \caption{Overview of the model. (a) Individual (word) placement in 2D space, (b) word list generation for the initial population, (c) movement and LLM-based competition, (d) payoff matrix reflecting the competitive relationship, and (e) mutation example showing possible word variations.}
  \label{fig:fig1}    
\end{figure*}

This study focuses on the ability of LLMs for generating diverse linguistic expressions to expand possible strategies or options in evolutionary game theory or possible properties of agents in agent-based evolutionary models. The evolutionary dynamics of societies based on interactions between individuals have long been discussed in evolutionary game theory. While it is possible to represent relationships between individuals with payoff matrices, the strategies available to each individual are mainly finite, and their relationship needs to be determined a-priori. On the other hand, in human and natural interactions between individuals, new options continue to emerge and evolve through lifetime and evolutionary time scale. 

Human language is characterized by complex relationships between words and their associated concepts, which can be evaluated from various perspectives. By incorporating the network of meanings and relationships inherent in words into agent-based models, we can realize more diverse and complex virtual ecological and evolutionary dynamics compared to conventional models.

This paper proposes a model for the evolutionary ecology of words that considers words (or phrases) as strategies or options in game-theoretical interactions. Specifically, we express the ecological and evolutionary dynamics created by a spatially interacting and evolving group of words by using LLMs for word creation, interaction judgment, and mutation. We introduce a minimal model and reports on preliminary experiments using the evolution of ``strong animal species" as an example scenario of evolution. 
Note that we aim to utilize the creativity of LLMs and their grasp of complex contextual word relationships to generate virtual evolutionary and ecological processes rather than simulating actual biological evolution.

With experiments with small populations, we demonstrate that from an initial population consisting of well-known animal species, many species emerged both gradually and in a punctuated equilibrium manner, including large terrestrial animals, marine life and extinct species, which are ecologically specialized and adapted ones across diverse extreme habitats. We also conducted a long-term experiment with a large population, demonstrating the emergence and coexistence of more diverse species. The codes and data are available online\footnote{\doi{10.6084/m9.figshare.28053425}}.

\section{Model}

\begin{figure*} % picture
    \centering
    \includegraphics[width=140mm]{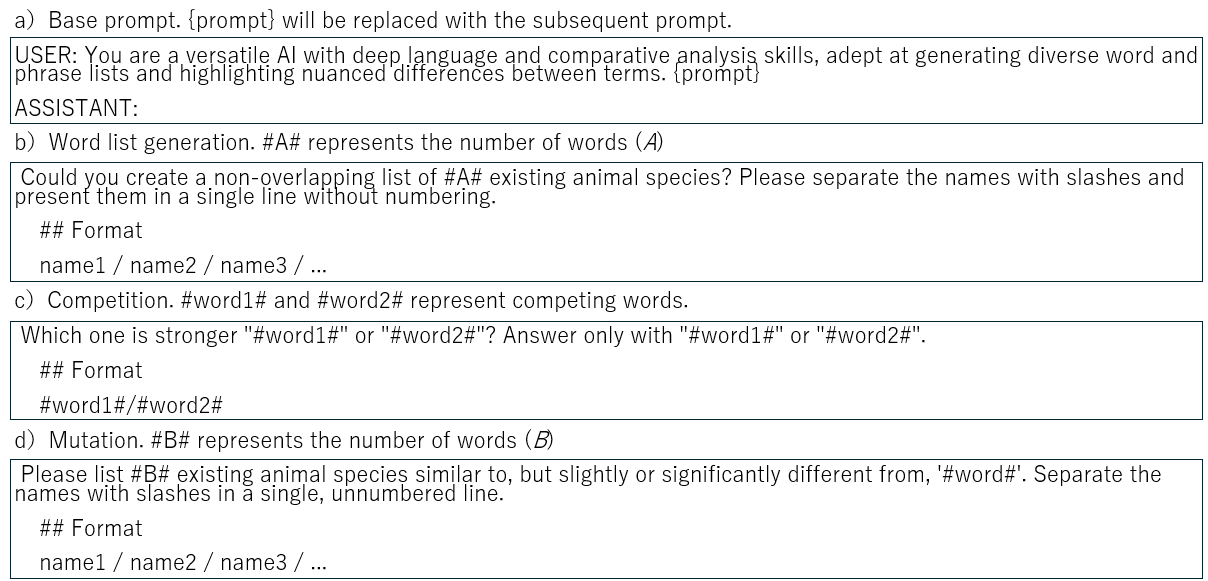}
  \caption{Prompts for ``strong animal species" survives. The base prompt (a), defines the basic behavior of the LLM, is prepended to each prompt for specific process: (b) initial word list generation, (c) competition judgment, and (d) mutation generation.}
  \label{fig:fig2}    
\end{figure*}

Figure \ref{fig:fig1} shows an overview of the model, and Figure \ref{fig:fig2} shows examples of prompts when using LLMs, adopting a setting where ``stronger animal species survives". As shown in Figure \ref{fig:fig1}a, $N$ individuals exist in a $W \times W$ two-dimensional toroidal grid space. Each individual has, in addition to its position information, a word expressed in English as a gene. A sequence of several words including spaces or short phrases are also allowed, but for simplicity, we will refer to them as words.

We use Gemma-2 \cite{Gemma:2024} developed by Google, an open-source chat-type LLM, in its quantized version (bartowski/gemma-2-9b-it-Q4\_K\_M.gguf on Hugging Face) for generation. First, we prepare a base prompt that gives characteristics to the LLM, as shown in Figure \ref{fig:fig1}a: ``You are a versatile AI with deep language and comparative analysis skills, adept at generating diverse word and phrase lists and highlighting nuanced differences between terms." We then add prompts appropriate for each usage scenario, as described below.

First, we ask the LLM to generate a list of $A$ words, as shown in Figure \ref{fig:fig2}b: ``Could you create a non-overlapping list of common names of \#A\# existing animals?" Each individual in the initial population holds a word randomly taken from this list and exists at a random position without overlap. Figure \ref{fig:fig1}b shows an example of a list.

At each step, the following is performed:
\begin{itemize}
    \item Movement: Each individual performs the following, in a random agent order. Each individual randomly selects a position from the adjacent 8 neighborhoods and moves there. However, if another individual already exists at the chosen position, it does not move. This setting is expected to create viscosity or gradual clustering as individuals are less likely to move when there are others in the neighborhood.
    
    \item Competition: Each individual performs the following, in a random agent order. If there are other individuals in the adjacent 8 neighborhoods, one is randomly selected as an opponent, and they compete with their own word against the opponent's word. The outcome is determined by asking the LLM with the prompt shown in Figure \ref{fig:fig2}c. The request is kept as simple as possible in this example, like ``Which one is stronger "\#word1\#" or "\#word2\#"?" without asking for reasons. Only if the LLM returns its own word is it judged to have won, and the losing individual's word is overwritten with its own word. Competitions between the same words are not conducted. Once a combination between words has been judged, the result is recorded in a dictionary, and subsequent competitions with the same combination refer to this to reduce computational cost. Figure \ref{fig:fig1}d shows an example of a competition between whale and eagle, where the LLM's response was whale. Note that while whales and eagles do not actually come into contact or compete in real-world ecosystems, whales are judged as stronger in the context of creative comparison, which may include their ecological properties, by LLMs we used. The payoff matrix in Figure \ref{fig:fig1}d shows this situation in which whale obtained $X$ and eagle obtained $Y$, and whale won the game because $X > Y$.
    
    \item Mutation: For each individual, its word mutates with probability $p_m$. The LLM is asked, as shown in Figure \ref{fig:fig2}d, ``Please list \#B\# existing animal species similar to, but slightly or significantly different from, '\#word\#'." to create a list of $B$ words. A word randomly selected from this list is adopted as the mutated word and replaces the existing one. This is to prevent the loss of diversity due to mutation, as simply requesting a single mutation word tends to generate the same mutation word for the same word. Figure \ref{fig:fig1}e shows an example of a whale mutating into a beluga whale.
\end{itemize}

It should be noted that the goal is not to simulate real biological evolution, but rather to utilize realistic and complex relationships among words in various contexts and the creativity of LLMs to generate virtual, complex evolutionary and ecological processes. Therefore, we can assume scenarios such as animals being judged as ``cute" will win, where the criteria may not necessarily be linked to ecological advantage. We can also assume scenarios where intangible elements, such as phrases, evolve based on more abstract criteria like wittiness.

\section{Preliminary Experiments}

\begin{figure*}[t] % picture
    \centering
    \includegraphics[width=140mm]{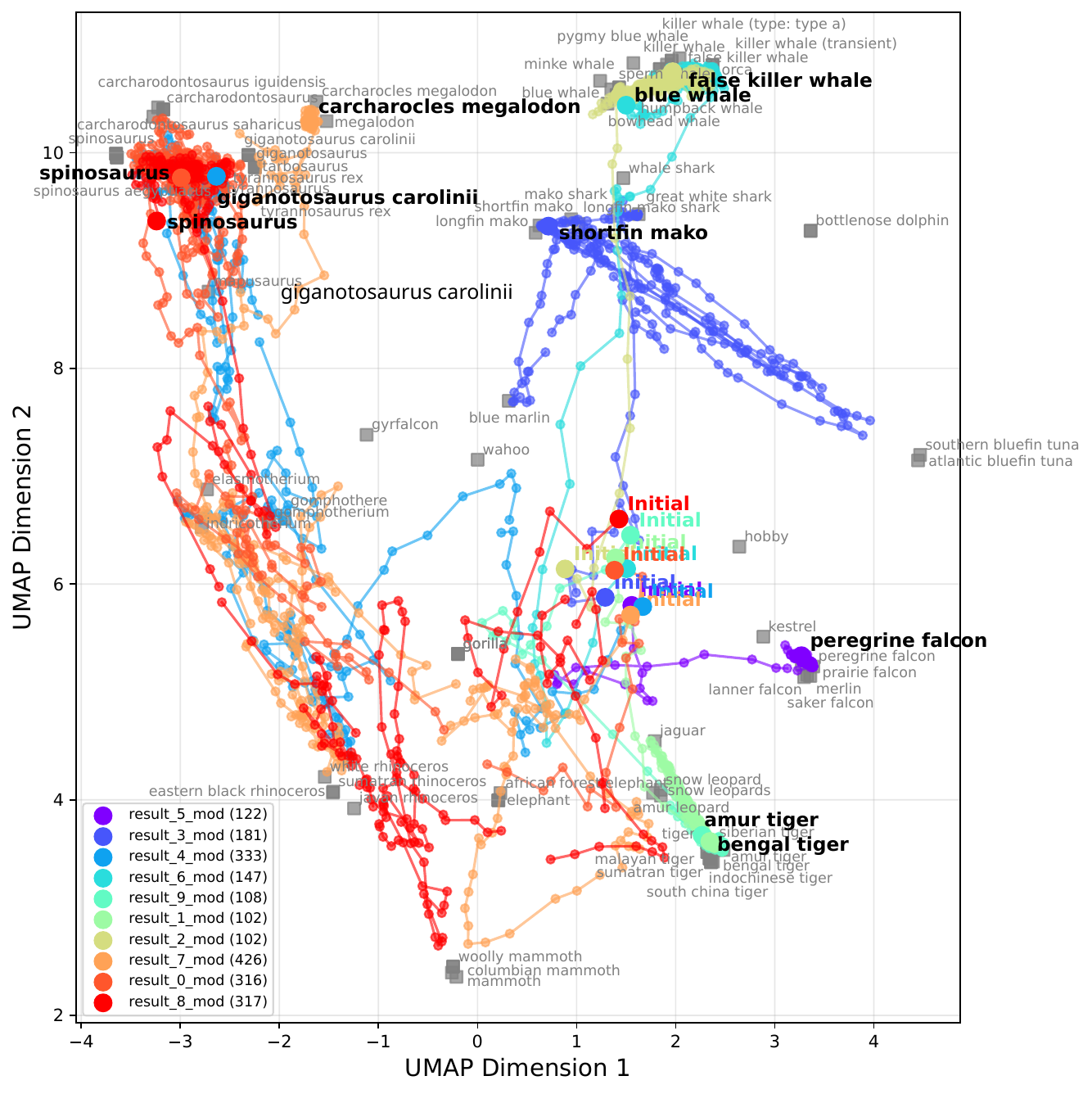}
  \caption{Transition of populations in the semantic space of words over 10 trials. Each dot represents the average word vector for each trial, plotted every 20 steps, with words vectorized in 2D using SentenceTransformer and UMAP. The 10 most frequent words in each trial (gray) and the most frequent word in the final step (bold black) are shown.}
  \label{fig:fig3}    
\end{figure*}

As a preliminary experiment, we conducted 10 trials of 300 steps each under the conditions of $W$=16, $N$=80, $p_m$=0.05, $A$=35, and $B$=10, using the prompts in Figure \ref{fig:fig3} assuming that ``strong animal species" survive.

Figure \ref{fig:fig3} shows the transition of the average word vector for each trial, plotted every 20 steps, with all words that appeared in all trials vectorized in 2D using SentenceTransformers \footnote{all-MiniLM-L6-v2} and UMAP \cite{McInnes:2018}, word vectorization and dimensionality reduction algorithms, respectively. The initial population is labeled as ``Initial", and the most frequent word in the final step is shown in large bold letters. The squares and the small words next to them show the words that ranked in the top 10 in frequency across all steps for each trial and their vectors.

\begin{figure*}[t] % picture
    \centering
    \includegraphics[width=130mm]{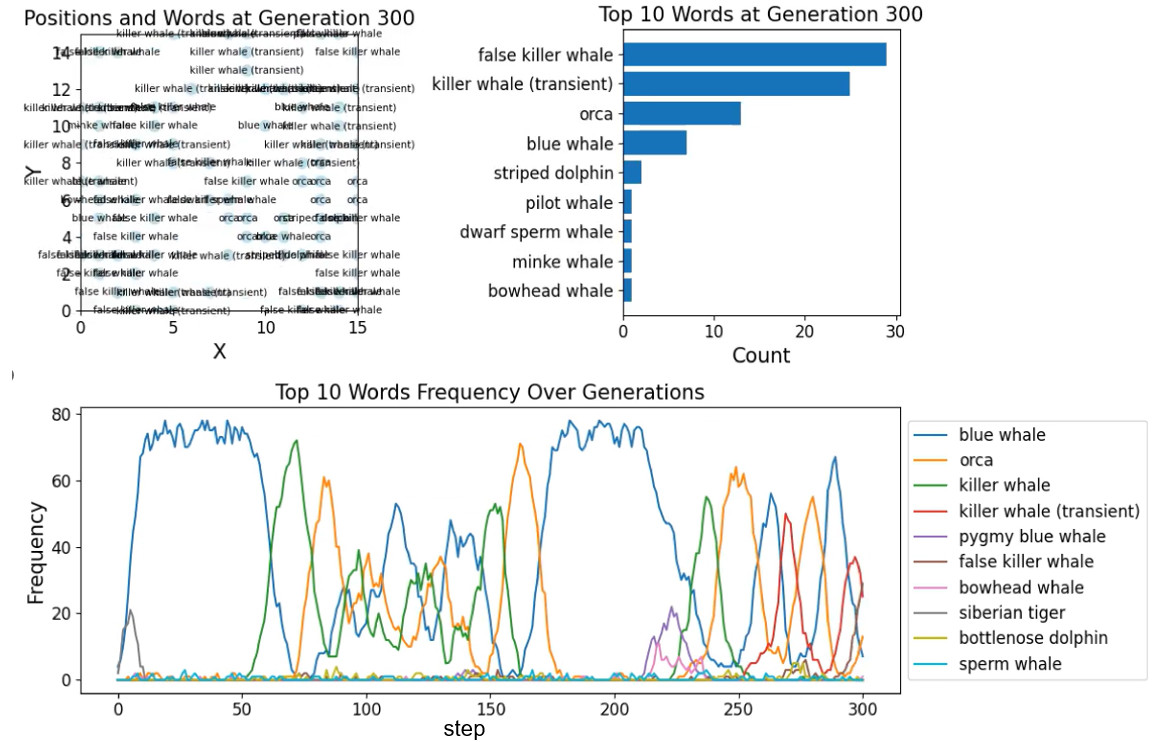}
  \caption{Spatial distribution of individuals in the final step of Trial 2 (top left), word frequency in the final step (top right), and transition of frequencies for the top 10 species through the trial (bottom).}
  \label{fig:fig4}    
\end{figure*}

From this figure, we can see that the initial populations in each trial are positioned similarly, slightly to the lower right of the center, indicating that they were generally composed of well-known animal species. Subsequently, diverse evolution was observed in each trial, with major species monopolizing or multiple related species frequently replacing each other. We found that there was a tendency for evolution towards mainly large animal species including terrestrial, marine, and extinct species. Focusing on the species that dominated the final step of each trial, for terrestrial animals, Amur tiger (One of the biggest cats, very strong hunter), Peregrine falcon (The fastest bird, can dive very fast to catch prey) and Bengal tiger (Strong and quick, top hunter in Indian forests) dominated at around the lower right in the figure. As for marine animals, Blue whale (The biggest animal in the world), False killer whale (Smart hunters that work together) and Shortfin mako (A very fast shark) dominated in the upper right. On the other hand, in the upper left, {\it Carcharocles megalodon} (The biggest shark ever, had scary big teeth), {\it Spinosaurus} (One of the biggest meat-eating dinosaurs, could live in water and on land), and {\it Giganotosaurus carolinii} (A big dinosaur as large as {\it T. rex}, had very strong jaws) became dominant. Until reaching these species, various species shown in gray emerged in the population, with over 100 species appearing in each trial, up to about more than 300 at most.

Figure \ref{fig:fig4} shows the frequency transition of the most frequent words in a typical trial (trial 2) along with the distribution of individuals and the list of top 10 frequent species in the final step. In this example, the population was dominated by blue whale until 50 and around 200 steps, followed by repeated processes of related marine animals becoming dominant in turns. As for the top 10 species, With the exception of the Siberian tiger, which is an apex predator on land and appeared at the initial stage of this experiment, they were marine species that have acquired various adaptive traits in the ocean: Blue whale (ocean giant, huge physical power), Orca (top predator, powerful hunter), Pygmy blue whale (smaller but strong sea animal), Bowhead whale (ice-breaking strength, long-lasting energy), Bottlenose dolphin (quick, smart hunter), and Sperm whale (strong deep-sea diver, strong ecolocator), Killer whale (same as orca, strong group hunter), Killer whale (transient) (special sea mammal hunter), and False killer whale (strong, social hunter), which dominated the population in the final step. 

It is interesting that the complex relationships between these species are not necessarily self-evident but were brought about by the output obtained from the LLM, yielding domination of a species and prey-predator dynamics. Although the population converged at a very early stage in some trials and others showed more dynamic transitions of dominant species, the emergence of major species occupying most of the population was repeatedly observed, as in this case.

\begin{figure*}[t] % picture
    \centering
    \includegraphics[width=140mm]{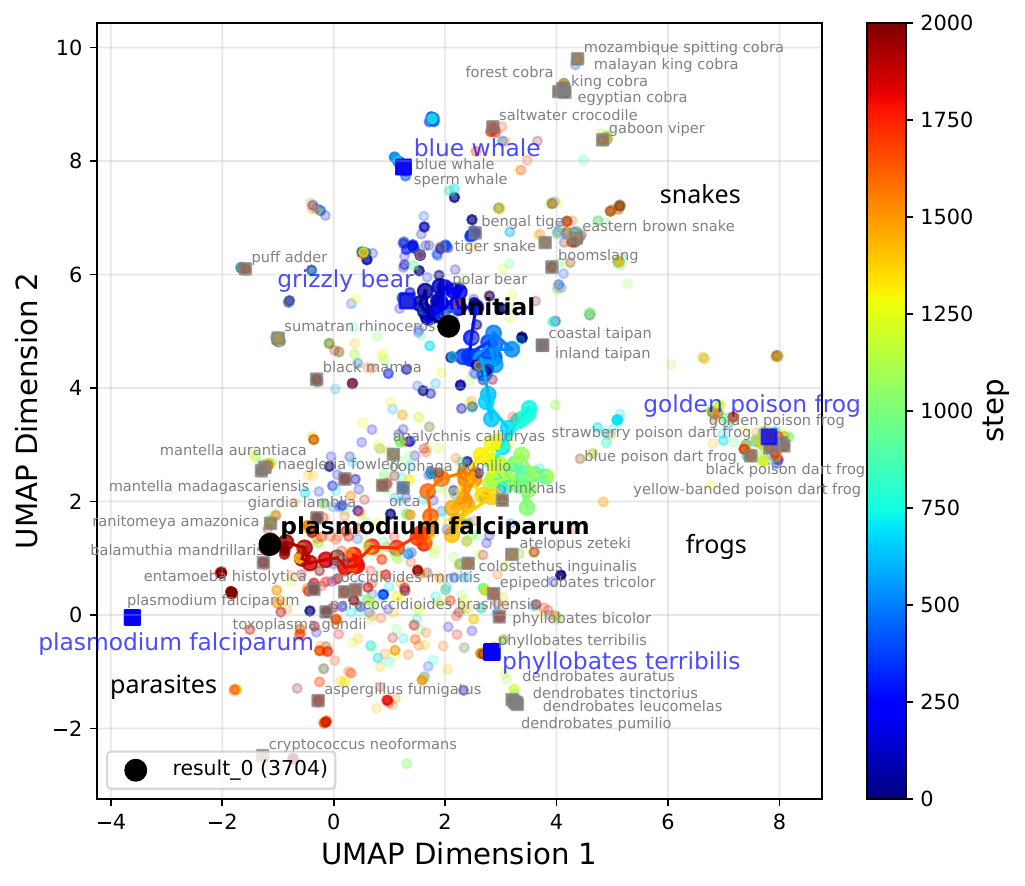}
  \caption{Transition of populations in the semantic space of words in a long-term trial. Each dot represents the average word vector plotted every 20 steps with color gradient (blue to red) indicating temporal progression. In order to illustrate the species diversity of the population, the word distributions are shown every 100 steps with the same color gradient indicating temporal progression. The most frequent word in the final step (bold black), 50 most frequent species (gray), and the dominant species every 100 steps (in blue) are shown.
  }

  \label{fig:fig5}    
\end{figure*}

Next, expecting the emergence of more diverse evolutionary processes, we conducted a longer-term (2000 steps) experiment using a larger population size ($N$=200) and two-dimensional space size ($W$=90). Figure \ref{fig:fig5} illustrates the temporal evolution of word vector averages sampled every 20 steps, along with the word distribution sampled every 100 steps (shown in different colors). Additionally, we plotted the distribution of the 50 most frequent species (in gray). We also plotted the dominant species every 100 steps (in blue), using their individual word vectors rather than population means and excluding duplicates.

The figure reveals several distinct phases in the evolutionary process. Initially, the population, originating near the center, was characterized by the dominance of blue whales, though they comprised only 20-30 \% of the population, with a slight movement toward the upper left continuing until approximately step 300. Subsequently, the population shifted rightward and downward, becoming predominantly composed of various poisonous species. These included the Inland Taipan (possessing the most potent venom among terrestrial snakes) and {\it Phyllobates terribilis} (Golden poison frog, recognized as Earth's most toxic vertebrate). The latter maintained dominance over 20-40 \% of the population until around step 1700. 

This period also saw the maintenance of diverse species, including mammals such as elephants and rhinoceroses, and extinct organisms like mammoths. The population occasionally included certain pathogenic fungi, notably {\it Batrachochytrium dendrobatidis} (Frog chytrid fungus), known for causing lethal skin infections in amphibians. In the final phase, the population shifted leftward, with {\it Plasmodium falciparum} (the most lethal malaria parasite) gradually becoming the dominant species, comprising up to 60\% of the population. Throughout the entire experiment, a total of 3,704 unique species names emerged.

As shown above, in this model where ``strong animal species" survive, we found that starting from the initial population composed of well-known animal species, the population evolved to ecologically specialized and adapted species across diverse extreme habitats and eras. It is interesting that extinct species and organisms like fungi and parasites, which were not initially considered, appeared and coexisted, emerging as novel options and introducing a new dimension to the evolutionary process.

\section{Conclusion}

In this paper, we proposed a word evolutionary ecology model that utilizes LLMs as a method to create finite options infinite in order to discuss open-ended evolution within the framework of agent-based models and evolutionary game theory. We first introduced experiments assuming a situation where ``strong animal species" survive in relatively small populations. The results showed that from the initial population, various species changed while replacing each other, evolving into diverse populations of terrestrial, marine, and extinct species. The common properties among them are that they are ecologically specialized and adapted species across diverse extreme habitats and eras, suggesting that diverse yet universal options emerged during the evolutionary process. 

In a longer-term experiment with a large population, various species coexisted and newly emerged in the population. Species that were substantially different from the initial population appeared and coexisted, which demonstrates important characteristics of LLMs as an engine for open-ended evolution, creating novelty and complexity. To examine such complex evolutionary dynamics, a challenge is to develop indices that measure characteristics of the evolving population, such as species diversity and adaptability, within this model.

On the other hand, in the proposed minimal model, the results of interactions (wins and losses) were decided deterministically by a single query to the LLM, which may have caused identical words to spread quickly. Therefore, for future work, we will consider introducing draws, probabilistic winner determinations, crossovers, and interactions among more individuals such as symbiotic relationships, to promote the maintenance of diversity. Additionally, describing local environmental conditions linguistically and assuming interactions among individuals under such environments can be considered as advantages of using LLMs to promote emergence and maintenance of words. Future work also includes experiments with more abstract evolutionary scenarios based on different word types like ``witty phrases survive" as explained before. 

Large language models like ChatGPT, which enable natural language interactions similar to humans, have attracted attention as attempts to understand and utilize them as autonomous agents. At the same time, we can extract relationships and structures in various worlds by setting up appropriate prompts. This research focuses on exploring the utilization of LLMs from this perspective, and we expect that it can contribute to the understanding and construction of open-ended artificial systems.

\section*{Acknowledgments}
We used Gemma-2 \cite{Gemma:2024} developed by Google, an open-source chat-type LLM, in its quantized version (bartowski/gemma-2-9b-it-Q4\_K\_M.gguf on Hugging Face) for generation in the proposed model. This study is supported in part by JSPS Topic-Setting Program to Advance Cutting-Edge Humanities and Social Sciences Research JPJS00122674991, JSPS KAKENHI 24K15103, and Google Gemma 2 Academic Program.
%\section*{References}

\end{document}